\documentclass[structabstract]{aa}  
\usepackage{graphicx}
\usepackage{txfonts}
\usepackage{natbib} 
\bibpunct{(}{)}{;}{a}{}{,}
\usepackage{lscape}
\begin{document}
   \title{Bimodality of light and s-elements in M4 (NGC~6121)
         \thanks{Based on observations made with ESO
         telescopes at La Silla Paranal Observatory under program ID 083.B-0083}.}

   \subtitle{A hint for the massive main-sequence star pollution scenario}

   \author{S. Villanova\inst{1}
          \and
          D. Geisler\inst{1}
          }

   \institute{Departamento de Astronomia, Casilla 160-C, 
              Universidad de Concepcion, Concepcion, Chile\\
              \email{svillanova-dgeisler@astro-udec.cl}
             }

   \date{Received ....; accepted ....}

 
  \abstract
   {All Globular Clusters (GCs) studied in detail so far  host two or more
    populations of stars (the multiple population phenomenon).
    Theoretical models suggest that the second population is formed from gas
    polluted by processed material produced by massive stars of the first
    generation. However the nature of the polluter is a matter of strong debate. Several
    candidates have been proposed: massive main-sequence stars (fast rotating or binaries),
    intermediate-mass AGB stars, or SNeII.
    }
   {We studied red giant branch (RGB) stars in the GC M4 (NGC~6121)
    to measure their chemical signature. Our goal is to measure
    abundances for many key elements (from Li to Eu) in order to give constraints
    about the polluters responsible for the multiple populations.
    }
   {We observed 23 RGB stars below the RGB-bump using the GIRAFFE@VLT2 spectroscopic facility.
    Spectra cover a wide range and allowed us to measure light
    (Li,C,$^{12}$C/$^{13}$C,N,O,Na,Mg,Al), $\alpha$ (Si,Ca,Ti,) , iron-peak
    (Cr,Fe,Ni), light-s (Y), heavy-s (Ba), and r (Eu) elements. We completed this study by analyzing a
    subsample of the UVES spectra presented in Marino et al. (2008) in order to have further
    clues about light s-elements of different atomic number (Y and Zr). 
    }
   {We confirm the presence of a bimodal population, first discovered by
    Marino et al. (2008). Stars can be easily
    separated according to their N content. The two groups have different
    C,$^{12}$C/$^{13}$C,N,O,Na content, but share the same
    Li,C+N+O,Mg,Al,Si,Ca,Ti,Cr,Fe,Ni,Zr,Ba and Eu abundance. Quite surprisingly the two
    groups differ also in their Y abundance. This result is strongly supported also
    by the analysis of the UVES spectra. 
   }
   {The absence of a spread in $\alpha$-elements, Eu and Ba makes SNeII and AGB stars unlikely
    as polluters. On the other hand, massive main-sequence stars can explain the
    bimodality of Y through the weak s-process. This stement is confirmed
    independently also by literature data on Rb and Pb.
    The lack of a Mg/Al spread and the extension of the [O/Na] distribution
    suggest that the mass of the polluters is between 20 and 30
    M$_{\odot}$. This implies a formation time scale for the cluster of 10$\div$30 Myrs. 
    This result is valid for M4. Other clusters like NGC~1851, M22, or
    $\omega$ Cen have different chemical signatures and may require other kinds
    of polluter.
    }


   \maketitle

\section{Introduction}

In the last few years, following the discovery of multiple populations in the 
color-magnitude diagrams (CMD) of some globular clusters (GCs) and in
spectroscopic samples of many of them, the debate on their
formation has been renewed. In this respect, the most interesting and peculiar 
clusters are  $\omega$ Centauri and NGC~2808, where at least 3 main sequences 
(MS) are present \citep{Be04, Vi07, Pi07}.

If those two objects represent the most extreme cases, it is now recognized that
all GCs studied in detail so far \citep{Ca09} show at least some kind of spread in their light element
content at the level of the RGB, the most evident being the spread in Na and O, elements that are
anti-correlated \citep{Ca10}. Na and O abundances are also (anti)correlated with other light
elements, such as C,N,Mg, and Al \citep{Gr04}.

\begin{figure}
\centering
\includegraphics[width=9cm]{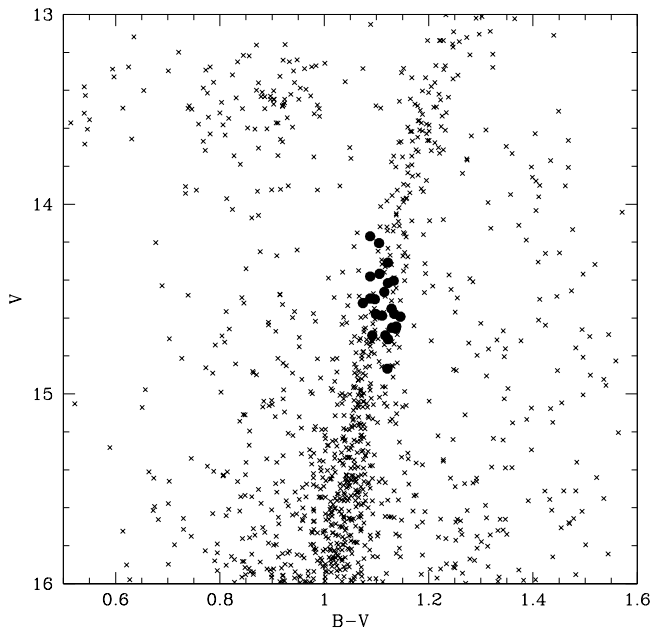}
\caption{The CMD of M4 with the observed RGB stars indicated as filled circles.}
\label{f1}
\end{figure}

The most natural explanation for this phenomenon is the self-pollution
scenario, where a cluster experiences an extended period of star formation, with the younger
population born from an interstellar medium polluted by ejecta
coming from stars of the older generation which have experienced hot H-burning
via p-capture. 
In this picture the older generation is the most He-N-Na-Al poor and C-O-Mg rich,
while the younger generation is affected by an enhancement of its He content,
together with N, Na, and Al, while C, O, and Mg turn out to be depleted.
This hypothesis can also explain correlations or anti-correlations 
of light elements present at the level of unevolved stars \citep{Gr01}.

Pollution must come from more massive stars. The main classes of
candidate polluters are:  intermediate-mass AGB stars
(\citealt{Ve02}, 4$<$M$<$7 M$_{\odot}$), fast-rotating massive main-sequence
(MS) stars (\citealt{De07}, M$>$15 M$_{\odot}$), and also massive MS binary
stars (\citealt{Mi09}, M$\sim$20 M$_{\odot}$).
All these channels can potentially pollute the existing interstellar material
with products of complete CNO cycle where N is produced at the expense of C and O,
the NeNa cycle, where Na is produced at the expense of Ne, and also the MgAl
cycle, where Al is produced at the expense of Mg (see \citealt{Re08} for an
extensive review).

AGBs eject part of their outer envelope during the thermal pulses after
they undergo hot-bottom burning, while massive stars eject material
through stellar winds contaminated by processed material brought to the surface
because of the fast rotation or binary interaction. In both cases the
primordial material (and the older generation) has the same composition as Galactic Halo
field stars (i.e. He-N-Na-Al poor and C-O-Mg rich), while the contaminated material (and the
younger generation) is He-N-Na-Al rich and C-O-Mg poor.
The material required to form the second generation is kept in the cluster due to the 
strong gravitational field \citep{De08}.

Another scenario was proposed by \citet{Mar09}. According to this paper
in a primordial metal-poor medium pre-polluted by SNe II explosion, AGB stars start to
eject their envelopes and a simultaneous SN Ia explosion further contaminates
(mainly with iron-peak elements) and collects the ejecta in a central
region. Here a first (older) generation is formed that, at
odds with the models described before, is He-N-Na-Al rich and C-O-Mg poor.
After that the most massive stars of this first generation evolve and explode
as SNeII, that pollute the remaining gas (mainly with $\alpha$-elements) and
mix it with the primordial medium. From this new material a second
(younger,  He-N-Na-Al poor and C-O-Mg rich) generation is formed. SNeII
produce also some amount of iron-peak elements but, due to this mix with the
primordial metal-poor medium, stars of the second generation have the same iron
content as the first generation. For our purposes the main point of
\citet{Mar09} is that SNeII are responsible for the chemical
inhomogeneities observed nowadays.

Also the abundance of other elements (including s- and r-process elements) 
may differ in stars of the first or second generation according to the nature
of the polluters.

While the pollution scenario is widely accepted nowadays, a key piece of
information is still missing.
We need to verify which kind(s) of polluter is responsible for the
contamination. This will help constrain the conditions and timescale of the
pollution process.

All processes described above produce (anti)correlations in light
elements (from C up to Al), but they behave
differently as far as other elements are concerned. AGB stars are known to produce
both light (i.e. Rb, Sr, Y, Zr) and heavy (i.e. Ba, La, Ce, Nd) s-elements
\citep{Bu01,Tr04} through the main-s process, while massive MS stars
(M$>$15M$_{\odot}$) produce only light s-element (up to A$\sim$90, i.e. up to
Y or Zr) through the weak-s process \citep{Ra93}.
Finally SNeII produce mainly $\alpha$-elements (e.g. Si and Ca) as well as r elements
(e.g. Eu), while SNeIa produce mainly iron-peak elements (e.g. Fe and Ni)
beside r elements \citep{Wa97}).
We can see that the study of $\alpha$, Fe-peak, light and heavy s, and r elements is
crucial to disentangle the proposed scenarios.

The aim of this paper is to measure abundances for a large sample of elements
in RGB stars of M4 (NGC~6121) in order to help constrain the nature of the
polluters. This cluster has been studied in detail
\citep{Iv99,Ma08,Yo08}. It has a bimodal Na-O distribution.
For this reason and because of its proximity, it is the ideal target for our
purposes. We would like to
verify which one of the proposed polluters, if any, can better explain the observed
abundance patterns. We will focus on a variety of s-elements, but we will include in our
analysis also many other elements (from Li up to Eu) in order to chemically
characterize the two sub-populations. 

In Section 2 we describe the observations. In Sec. 3 and 4 we discuss the
determination of the abundances and present the results. In Sec. 5 we discuss
the results, while Sec. 6 gives the conclusions.

\begin{figure*}
\centering
\includegraphics[width=9cm]{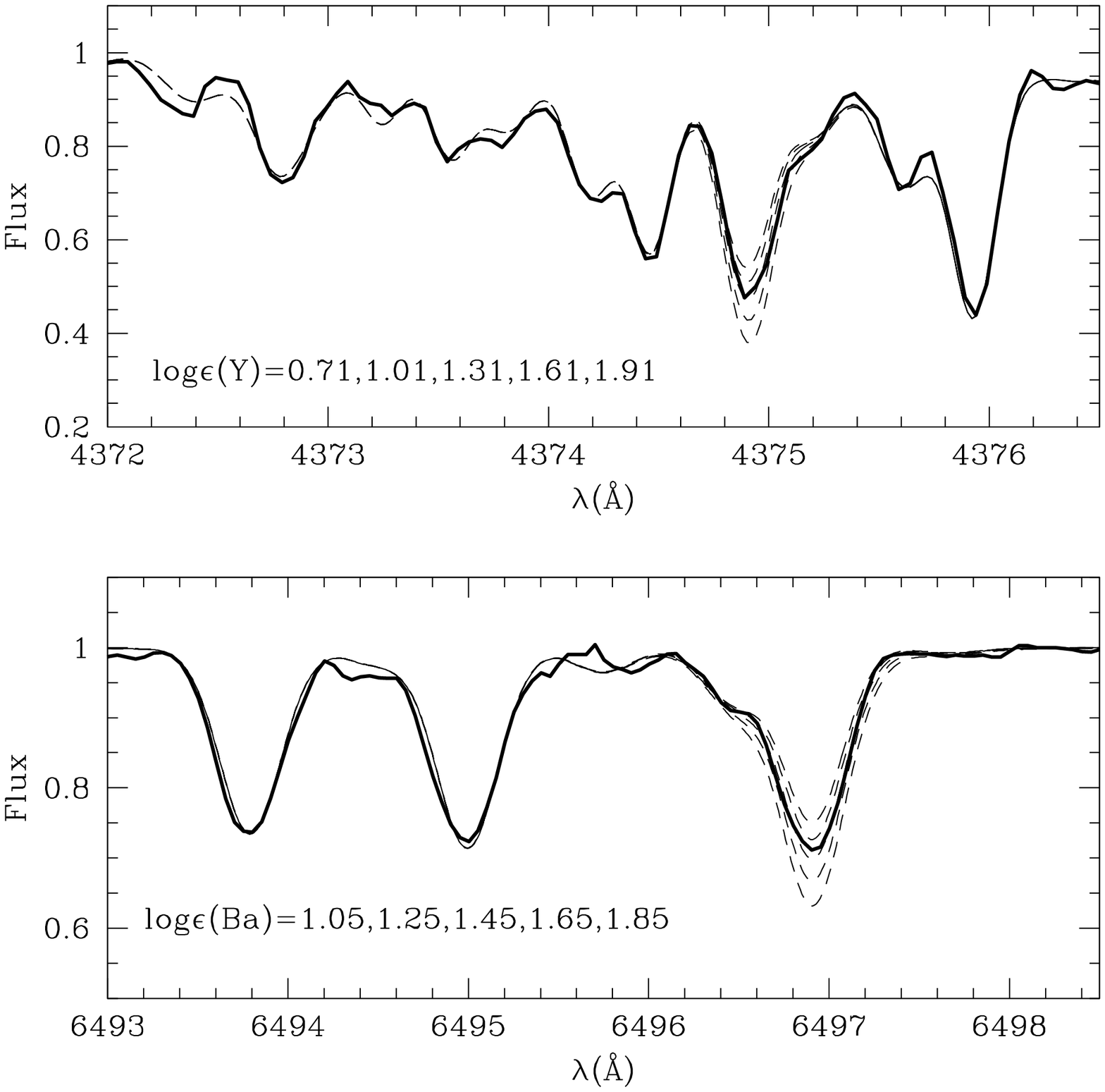}
\includegraphics[width=9cm]{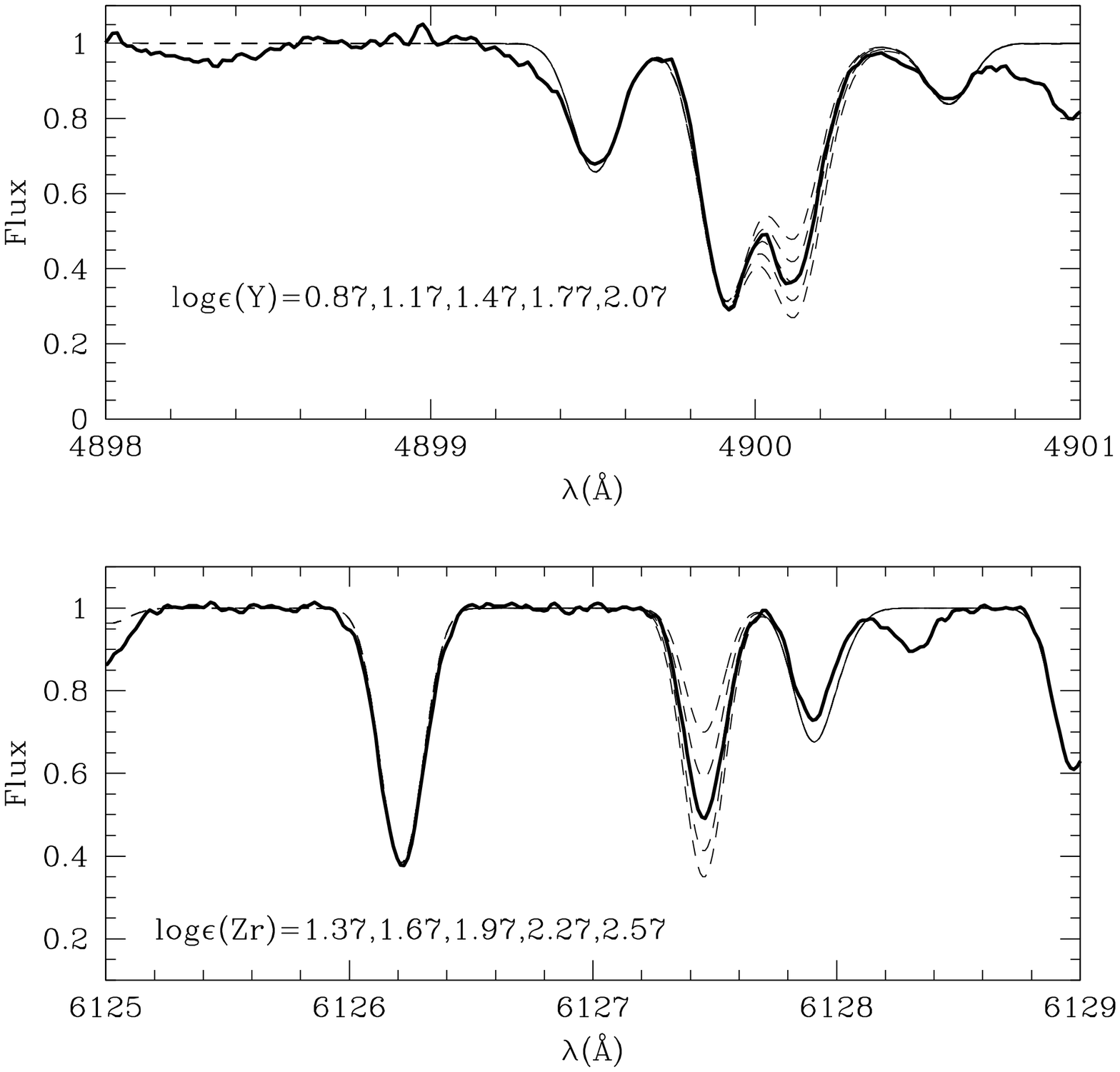}
\caption{Example of spectral synthesis for Y and Ba lines applied to GIRAFFE
data (left panels, star \#907), and to UVES data (right panel, star \#19925).
Abundances used in the spectral synthesis are indicated.}
\label{f2}
\end{figure*}

\section{Observations and data reduction}

Our dataset consists of high resolution spectra collected in June-August 2010.
The spectra come from 10$\times$45m exposures, obtained with the FLAMES-GIRAFFE
spectrograph, mounted at the VLT telescope.
Weather conditions were good with a typical seeing of $\sim$1.0 arcsec. 
We selected 23 isolated stars at V$\sim$ 14.5, located below the RGB-bump
of the cluster, from B,V photometry (\citealt{Mo03}, see
Fig.~\ref{f1}). 
All the stars lie within $\pm$0.3 mag, so
can be assumed to be in the same evolutionary phase.
All stars were observed with 4 different set-ups, HR04 (4 exposures, range= 4188-4392 \AA,R=20000), 
HR11 (2 exposures, range= 5597-5840 \AA, R=24000), 
HR13 (2 exposures, range=6120-6405 \AA, R=22000), and
HR15N (2 exposures, range= 6470-6790 \AA, R=17000).
 
Data were reduced using the dedicated pipeline BLDRS v0.5.3, written at the
Geneva Observatory (see http://girbldrs.sourceforge.net).
Data reduction includes bias subtraction, flat-field correction, wavelength calibration,
sky subtraction, and spectral rectification. Spectra have a typical S/N of
$\sim$150 at 6300 \AA.\\

Radial velocities were measured by the {\it fxcor} package in IRAF,
using a synthetic spectrum as a template. 
The mean heliocentric value for our targets is 71.9$\pm$0.9 km/s, while
the dispersion is 4.2$\pm$0.6 km/s. \citet{So09} gives
70.3$\pm$0.2 km/s as heliocentric radial velocity for M4, 
and the typical dispersion for a cluster of its mass is $\sim$4-5 km/s
\citep{Pr93}, in very good agreement with our results.
On the basis of this result, we conclude that all our targets 
are cluster members.

Table~\ref{t1} lists the basic parameters of the selected stars:
the ID, the J2000.0 coordinates (RA \& DEC), U,B,V magnitudes
(Momany, private communication), heliocentric radial velocity (RV$_{\rm H}$),
T$_{\rm {eff}}$, log(g), micro-turbulence velocity (v$_{\rm t}$). For
determination of atmospheric parameters see the next section.

\begin{table*}
\caption{Basic parameters of the observed stars.}            
\label{t1}      
\centering                          
\begin{tabular}{lccccccccc}        
\hline\hline                 
ID & RA(degrees) & DEC(degrees) & U(mag) & B(mag) & V(mag) &
RV$_{\rm H}$(km/s) & T$_{\rm {eff}}$(K) & log(g)(dex) & v$_{\rm t}$(km/s)\\    
\hline             
28590 & 245.81437500 & -26.64647222 & 16.13 & 15.60 & 14.52 & 73.8 & 4920 & 2.65 & 1.07\\
33584 & 245.79137500 & -26.47600000 & 16.14 & 15.60 & 14.50 & 77.0 & 4940 & 2.67 & 1.07\\
36820 & 245.98212500 & -26.65019444 & 16.33 & 15.81 & 14.69 & 71.0 & 4940 & 2.73 & 1.12\\
37614 & 245.93579167 & -26.63430556 & 16.13 & 15.68 & 14.55 & 73.9 & 4940 & 2.55 & 1.10\\
39100 & 245.95291667 & -26.60716667 & 16.04 & 15.54 & 14.40 & 70.7 & 4890 & 2.68 & 1.00\\
40197 & 245.93508333 & -26.59366667 & 16.33 & 15.78 & 14.65 & 79.6 & 4940 & 2.75 & 1.16\\
41863 & 245.86783333 & -26.57441667 & 16.40 & 15.83 & 14.71 & 67.4 & 4940 & 2.65 & 1.24\\
42561 & 245.95445833 & -26.56680556 & 16.22 & 15.70 & 14.59 & 66.4 & 4950 & 2.77 & 0.94\\
43020 & 245.87895833 & -26.56230556 & 16.48 & 15.99 & 14.87 & 76.1 & 5030 & 3.05 & 1.08\\
43085 & 245.86550000 & -26.56161111 & 16.33 & 15.74 & 14.59 & 76.4 & 5000 & 2.95 & 1.13\\
43494 & 245.87850000 & -26.55780556 & 16.28 & 15.79 & 14.66 & 65.8 & 4970 & 2.67 & 1.20\\
43663 & 245.84891667 & -26.55608333 & 16.23 & 15.68 & 14.58 & 78.0 & 4960 & 2.90 & 1.00\\
45171 & 245.86562500 & -26.54222222 & 16.26 & 15.71 & 14.58 & 77.6 & 4930 & 2.90 & 0.88\\
45200 & 245.92258333 & -26.54197222 & 16.36 & 15.78 & 14.65 & 67.4 & 5010 & 2.70 & 1.17\\
46201 & 245.93333333 & -26.53383333 & 15.98 & 15.47 & 14.38 & 70.6 & 4930 & 2.45 & 1.08\\
47596 & 245.86287500 & -26.52325000 & 16.07 & 15.58 & 14.46 & 68.4 & 4960 & 2.70 & 1.23\\
48499 & 245.95400000 & -26.51641667 & 15.92 & 15.43 & 14.31 & 76.0 & 4960 & 2.85 & 1.10\\
49381 & 245.88962500 & -26.50972222 & 16.26 & 15.78 & 14.69 & 69.8 & 4940 & 2.55 & 1.22\\
50032 & 245.93904167 & -26.50455556 & 16.13 & 15.59 & 14.50 & 74.0 & 5050 & 3.03 & 0.99\\
53602 & 245.83920833 & -26.47597222 & 16.17 & 15.54 & 14.42 & 68.0 & 5000 & 3.05 & 1.00\\
67553 & 246.00250000 & -26.47241667 & 15.77 & 15.26 & 14.17 & 66.6 & 4900 & 2.60 & 1.02\\
8460  & 245.94720833 & -26.37500000 & 16.14 & 15.47 & 14.37 & 70.4 & 4930 & 2.50 & 1.18\\
907   & 246.03741667 & -26.37791667 & 15.81 & 15.31 & 14.21 & 70.1 & 4920 & 2.65 & 1.13\\
\hline                                   
\end{tabular}
\end{table*}

\begin{figure*}[ht]
\centering
\includegraphics[width=15cm]{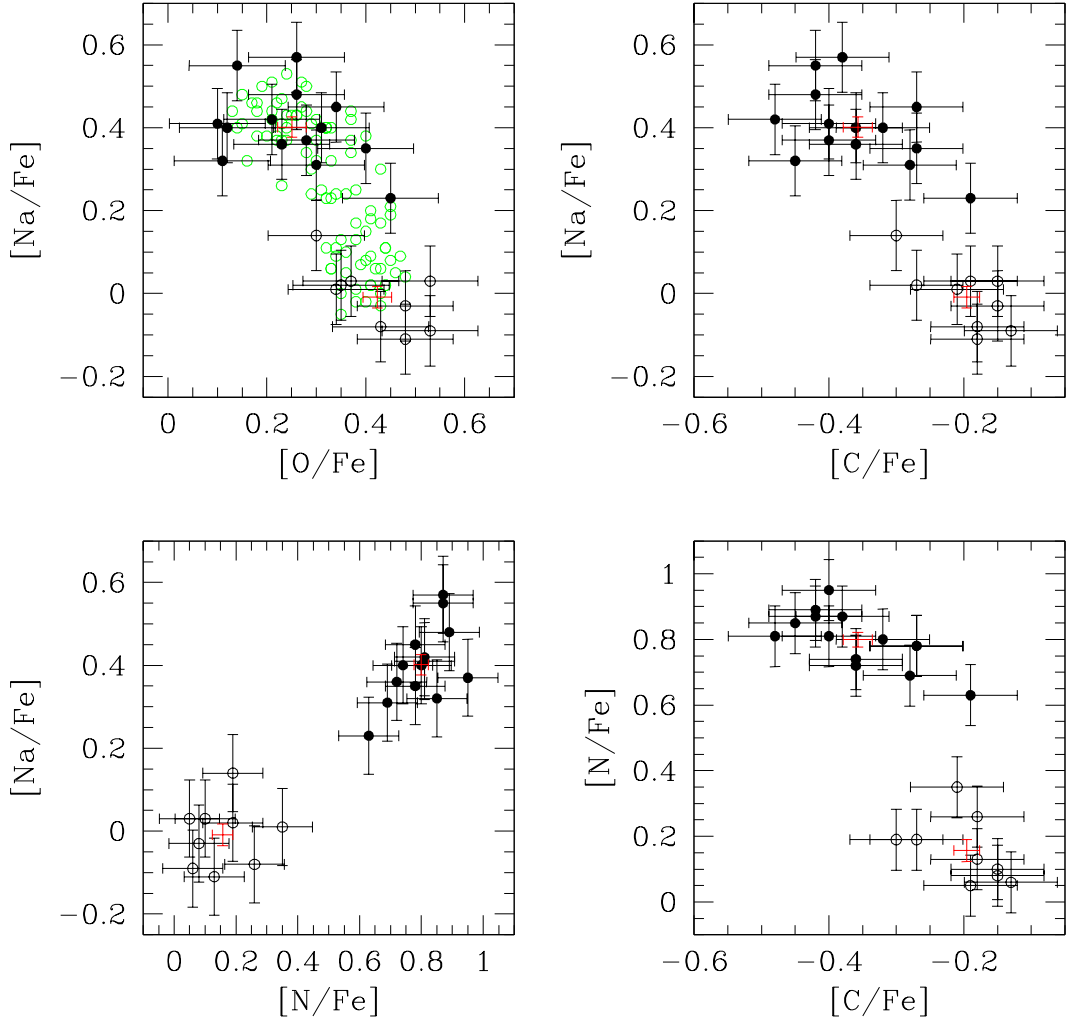}
\caption{Correlation between light-elements (C,N,O,Na) as obtained from our
GIRAFFE stars. Green points in the upper left panel are the results by
\citet{Ma08}. N-poor stars are indicated as open circles, while N-rich as
filled circles. Red crosses represent the mean value for each sub-population.}
\label{f3}
\end{figure*}

\section{Abundance analysis}

The chemical abundances for Na, Mg, Al, Si, Ca, Ti, Cr, Fe, and Ni were
obtained from the equivalent widths (EWs) of the spectral lines. 
See \citet{Ma08} for a more detailed explanation of the method we
used to measure the EWs.
For the other elements (Li, C, N, O, $^{12}$C/$^{13}$C, Y, Ba, Eu), whose lines
are affected by blending, we used the spectrum-synthesis method. For this
purpose we calculated 5 synthetic spectra having different abundances for the
element, and estimated the best-fitting value as the one that minimize the r.m.s.
Na and Al present few features in the spectrum, so in this case abundances derived from
the EWs were cross-checked with the spectral synthesis method in order to
obtain more accurate measurements. 
Only lines not contaminated by telluric lines were used. 

Atmospheric parameters were obtained in the following way. 
First of all T$_{\rm eff}$ was derived from the B-V color using the relation by 
\citet{Al99} and the reddening (E(B-V)=0.36) from \citet{Ha96}. 
Surface gravities (${\log(g)}$) were obtained from the canonical equation:

\begin{displaymath}
\log(\frac{g}{g_{\odot}}) = \log(\frac{M}{M_{\odot}}) + 4\cdot
\log(\frac{T_{\rm eff}}{T_{\odot}}) - \log(\frac{L}{L_{\odot}})
\end{displaymath}

where the mass ${M/M_{\odot}}$ was assumed to be 0.8 M$_{\odot}$, and the
luminosity ${L/L_{\odot}}$ was obtained from the absolute magnitude ${M_{\rm V}}$
assuming an apparent distance modulus of ${(m-M)_{\rm V}}$=12.82 \citep{Ha96}. The
bolometric correction (BC) was derived by adopting the relation 
BC-T$_{\rm eff}$ from \citet{Al99}.
Finally, micro-turbulence velocity (${v_{\rm t}}$) was obtained from the
relation of \citet{Ma08}.\\
These atmospheric parameters were considered as initial estimates and were refined during the
abundance analysis. As a first step atmospheric models were calculated using ATLAS9 \citep{Ku70}
and assuming the initial estimate of T$_{\rm eff}$, ${\log(g)}$,
and ${v_{\rm t}}$, and the [Fe/H] value from \citet{Ha96}.\\ 
Then T$_{\rm eff}$, ${v_{\rm t}}$, and ${\log(g)}$ were adjusted and new
atmospheric models calculated in an interactive way in order to remove trends in
Excitation Potential (E.P.) and equivalent widths vs. abundance for 
${T_{\rm eff}}$ and ${v_{\rm t}}$ respectively, and to satisfy the ionization
equilibrium for ${\log(g)}$. FeI and FeII were used for this purpose. 
The [Fe/H] value of the model was changed at each iteration according to the
output of the abundance analysis.
The Local Thermodynamic Equilibrium (LTE) program MOOG \citep{Sn73} was used
for the abundance analysis.

We checked the reliability of our atmospheric parameters by comparing
photometric and spectroscopic T$_{\rm eff}$, 
finding a mean difference in temperature lower than 50 K.\\ 
A further check was performed on our log(g) scale. We inverted the
previous equation in order to obtain the mass and calculated the mean mass of
our targets. We got $\overline{M}$$=0.83\pm0.06\ M_{\odot}$, in good agreement with the
value obtained from isochrone fitting ($\sim$0.8 M$_{\odot}$).
Our conclusion is that our T$_{\rm eff}$ and log(g) values can be safely used to
obtain abundances.

The linelists for the chemical analysis were obtained from many sources
(\citet{Gr03}, VALD \& NIST\footnote{See
http://vald.astro.univie.ac.at/$\sim$vald/php/vald.php and http://physics.nist.gov/PhysRefData/ASD/lines\_form.html}, \citet{Mw94},
\citet{Mw98}, SPECTRUM\footnote{See   http://www.phys.appstate.edu/spectrum/spectrum.html 
and references therein}, and SCAN\footnote{See http://www.astro.ku.dk/$\sim$uffegj/}), 
and calibrated using the Solar-inverse technique by the spectral synthesis
method (see \citealt{Vi09} for more details). For this purpose we used the high
resolution, high S/N NOAO Solar spectrum \citep{Ku84}. 
Adopted solar abundances we obtained with our linelist are reported in
Tab.~\ref{t3} and \ref{t4} together with those ones by \citet{Gr98} for comparison.
We emphasize the fact that all the linelists were calibrated on the Sun,
including those used for the spectral synthesis. 

Lines treated with the EQW method are reported in Tab.~\ref{t6}
(electronic edition) together with the adopted parameters and equivalent
widths star-by-star. Parameters for lines treated with the spectral synthesis
method are not reported because the line list would be too long (thousands
of lines in some cases). For these lines references are given above.

Li was measured from the line at 6707 \AA, while our determinations of 
C,N,O abundances are based on the G-band at 4310 \AA, the CN band at 4215 \AA, and the
forbidden O line at 6300 \AA\ respectively. CN lines at 4230 \AA\ were used also
to estimate the $^{12}$C/$^{13}$C ratio.

These features were also checked on the high resolution, high S/N
spectrum of Arcturus. See \citet{Vi10} for more details.
Abundances for C, N, and O were determined all together in an
interactive way in order to take into account any possible molecular coupling 
of these three elements.

Our targets are objects evolved off the main sequence, so some evolutionary
mixing is expected. This can affect the primordial C,N,O abundances separately,
but not the total C+N+O content because these elements are transformed one
into the other during the CNO cycle. Is does not affect the relative C,N,O of
our stars either, since the stars are all in the same evolutionary phase.

Na was obtained from lines at 6154 and 6160 \AA\ (EQW) and 5682 and 5688 \AA,
and corrected for NLTE effects following the prescription by \citet{Gr99}.
Mg was obtained from the line at 5711 \AA\ while Al from the lines at 6696 and 6699 \AA.
Y abundance was measured using the line at 4375 \AA\ while for Ba we used the
line at 6494 \AA (see Fig.~\ref{f2}, left panels). For Ba we took the hyperfine splitting into account
using \citet{Mw94} and \citet{Mw98} data.
Finally Eu was obtained from the line at 6645 \AA.

In order to extend and confirm our results, we analyzed also a subsample
of 24 stars of \citet{Ma08} observed with UVES, the high-resolution
spectrograph mounted at VLT telescope. As parameters we used those published
there, but we extended the chemical analysis to Y and Zr (see Fig.~\ref{f2},
right panels), two elements not considered in that paper. Y was obtained from
the line at 4900 \AA, while Zr from the line at 6127 \AA. Also in this case
abundances were obtained by spectrum-synthesis. In UVES spectra two other
Y lines were available, at 4883 and 5087 \AA, but the one at 4900 \AA\ turned
out to be stronger and less affected by noise. It is partially blended with
another line as is visible in Fig.~\ref{f2}, but this can be easily
managed by the spectrum-synthesis method we applied.

An internal error analysis was performed by varing T$_{\rm eff}$, log(g), [Fe/H], and
v$_{\rm t}$ and redetermining abundances of star \#33584, assumed to represent
the entire sample. Parameters were varied by $\Delta$T$_{\rm eff}$=+50 K,
$\Delta$log(g)=+0.10, $\Delta$[Fe/H]=+0.05 dex, and $\Delta$v$_{\rm t}$=+0.1
km/s. This estimation of the internal errors for atmospheric parameters was
performed as in \citet{Ma08}.
Results are shown in Tab.~\ref{t5}, including the error due to the noise
of the spectra. This error was obtained, for elements whose abundance was
obtained by EQWs, as the average value of the errors on the mean as given by
MOOG, and for elements whose abundance was obtained by spectrum-synthesis, as
the error given by the fitting procedure. $\sigma_{\rm tot}$ is the
squared sum of the single errors, while $\sigma_{\rm obs}$ is the mean
observed dispersion of the two sub-populations of the clusters (as identified
by their N content, see next section). The agreement between the two values is
reasonable for all elements.

\section{Results}

First of all, in Fig.~\ref{f3} we plot the abundance of C,N,O,Na.
In all the panels we clearly see that the distribution is bimodal for all the
four elements considered. Na-O anti-correlation is compared with \citet{Ma08}
(green points). We confirm that stars in the cluster are divided in two well
separated groups having different light-element content. 
In the following analysis we take as reference the lower left panel of
Fig.~\ref{f3} because N appears to be the best element to separate the two
groups. From now on in all the figures and in the discussion of the results, we
divide our stars into N-poor (open circles in all the figures), 
and N-rich (filled  circles in all the figures). 

We note that in our sample 9 stars belong to the N-poor
group (the so called first generation), while 14 to the N-rich  group (the so
called second generation). This means that about 40\% of our stars belong to
the first generation. This agree within the errors ($\sigma\sim\pm$6\%) with \citet{Ca09} and
\citet{Ma08}, where the authors find that the cluster is composed of $\sim$30\% and
$\sim$50\%, respectively, first generation stars.

Errors on the single measurements (the black errorbars in the
figures) for a given element are the $\sigma_{\rm obs}$ of Tab.~\ref{t5}.
The red crosses in the figures represent the mean abundance and the 
error of the mean for each group.
We see that N-poor stars are also C-rich, O-rich, and Na-poor, while N-rich
stars are also C-poor, O-poor, and Na-rich, in accord with the theoretical
expectations outlined previously.

Mean abundances we obtained for the two groups and for the cluster are summarized in
Tab.~\ref{t2}, while abundances for each star are summarized in
Tab.~\ref{t3}, and \ref{t4}. 

With respect to the mean abundances of the two groups, for each element in the 4$^{th}$ colum of
Tab.~\ref{t2} we report the abundance difference (i.e. the significance) in units of $\sigma_{el}$
that is defined as:

\begin{displaymath}
\sigma_{el}=\sqrt{\sigma_{el,N-poor}^{2}+\sigma_{el,N-rich}^{2}}
\end{displaymath}

where $\sigma_{el,N-poor}$ and $\sigma_{el,N-rich}$ are the errors on the mean
abundance of the two groups as given by the 2$^{nd}$ and 3$^{rd}$ columns of Tab.~\ref{t2}.
This value tell us if this difference is significant with a value of
$\sigma_{el}>$3 implying strong significance.
The second part of Tab.~\ref{t4} reports the results obtained from the analysis
of the UVES spectra of \citet{Ma08}. Na abundances are obtained from that
paper. For these stars we do not have the N content. However it is clear from
Fig.~\ref{f3} that N-poor and N-rich stars can be easily identified also by their Na
content. So UVES stars were classified according to their [Na/Fe] value.
All stars with [Na/Fe]$<$0.23 were considered N-poor, all stars with
[Na/Fe]$\geq$0.23 were considered N-rich.

\begin{table*}
\caption{Mean abundances of the two groups of stars (2$^{nd}$ and 3$^{rd}$
column). The 4$^{th}$ column is the the significance of the abundance
difference of the two groups, in units if $\sigma_{el}$. 5$^{th}$ column gives the
mean abundance of the cluster as the average of the two groups. 6$^{th}$,
7$^{th}$, and 8$^{th}$ columns are the abundances by \citet[Ma08]{Ma08},
\citet[Iv99]{Iv99}, and \citet{Yo08}+\citet{Yo08b}=Yo08, respectively. The results
of the last two papers are reported together because they are complementary.}
\label{t2}      
\centering                          
\begin{tabular}{lccccccc}        
\hline\hline                 
El. & N-poor & N-rich & Sig. (units of $\sigma_{el}$) & M4(this work) & Ma08 & Iv99 & Yo08\\
\hline   
\multicolumn{8}{c}{GIRAFFE data}\\
\hline
log$\epsilon$(Li)       & +0.97$\pm$0.04 & +0.97$\pm$0.03 & 0.0  & +0.97 &    -  &    -  &    -  \\
${\rm [C/Fe]}$          & -0.20$\pm$0.02 & -0.36$\pm$0.02 & 5.7  & -0.28 &    -  & -0.50 &    -  \\
${\rm [N/Fe]}$          & +0.16$\pm$0.03 & +0.80$\pm$0.02 & 17.8 & +0.48 &    -  & +0.85 &    -  \\
${\rm [O/Fe]}$          & +0.42$\pm$0.03 & +0.25$\pm$0.03 & 4.0  & +0.34 & +0.39 & +0.25 & +0.56 \\
log$\epsilon$(C+N+O)    &  8.18$\pm$0.03 &  8.14$\pm$0.02 & 1.1  &  8.16 &    -  &  8.24 &    -  \\
$^{12}$C/$^{13}$C       & 21.7$\pm$0.8   & 17.4$\pm$1.0   & 3.4  & 19.6  &    -  &   4.5 &    -  \\
${\rm [Na/Fe]}$         & -0.01$\pm$0.03 & +0.40$\pm$0.02 & 11.4 & +0.20 & +0.27 & +0.22 & +0.43 \\
${\rm [Mg/Fe]}$         & +0.46$\pm$0.03 & +0.48$\pm$0.02 &  0.5 & +0.47 & +0.50 & +0.44 & +0.57 \\
${\rm [Al/Fe]}$         & +0.51$\pm$0.04 & +0.53$\pm$0.02 & 0.4  & +0.52 & +0.54 & +0.64 & +0.74 \\
${\rm [Si/Fe]}$         & +0.43$\pm$0.02 & +0.42$\pm$0.02 & 0.4  & +0.43 & +0.48 & +0.65 & +0.58 \\
${\rm [Ca/Fe]}$         & +0.42$\pm$0.01 & +0.40$\pm$0.02 & 0.9  & +0.41 & +0.28 & +0.26 & +0.42 \\
${\rm [Ti/Fe]}$         & +0.35$\pm$0.02 & +0.31$\pm$0.01 & 1.8  & +0.33 & +0.32 & +0.30 & +0.41 \\
${\rm [Cr/Fe]}$         & +0.00$\pm$0.02 & +0.01$\pm$0.03 & 0.3  & +0.01 & -0.04 &    -  & +0.08 \\
${\rm [Fe/H]}$          & -1.14$\pm$0.01 & -1.14$\pm$0.02 & 0.0  & -1.14 & -1.07 & -1.18 & -1.23 \\
${\rm [Ni/Fe]}$         & +0.00$\pm$0.01 & -0.02$\pm$0.01 & 1.4  & -0.01 & +0.02 & +0.05 & +0.12 \\
${\rm [Y/Fe]}$          & +0.10$\pm$0.06 & +0.31$\pm$0.03 & 3.1  & +0.21 &    -  &    -  & +0.69 \\
${\rm [Ba/Fe]}$         & +0.29$\pm$0.02 & +0.32$\pm$0.01 & 1.3  & +0.31 & +0.41 & +0.60 &    -  \\
${\rm [Eu/Fe]}$         & +0.20$\pm$0.03 & +0.20$\pm$0.03 & 0.0  & +0.20 &    -  & +0.35 & +0.40 \\
\hline           
\multicolumn{8}{c}{UVES data}\\
\hline
${\rm [Na/Fe]}$         & +0.07$\pm$0.03 & +0.42$\pm$0.02 & 9.7  & +0.25 & +0.27 & +0.22 & +0.43 \\
${\rm [Y/Fe]}$          & +0.19$\pm$0.03 & +0.33$\pm$0.02 & 3.9  & +0.26 &    -  &    -  & +0.69 \\
${\rm [Zr/Fe]}$         & +0.44$\pm$0.02 & +0.47$\pm$0.03 & 0.8  & +0.46 &    -  &    -  & +0.23 \\
\hline
\end{tabular}
\end{table*}

\begin{figure}
\centering
\includegraphics[width=9cm]{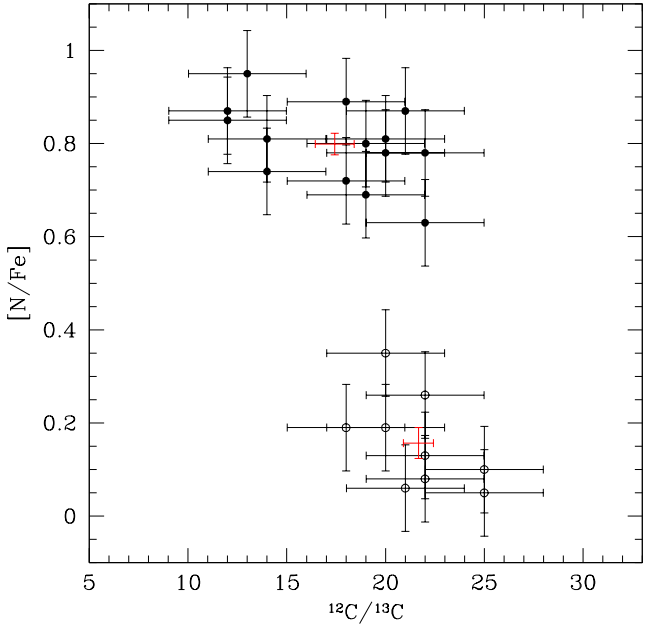}
\caption{[N/Fe]-$^{12}$C/$^{13}$C anti-correlation for our GIRAFFE stars.
Red crosses represent the mean value for each sub-population.}
\label{f4}
\end{figure}

Comparing values in Tab.~\ref{t2} we see immediately that the two
sub-populations have the same content (difference of 1.8$\sigma$ in the worst
case) of $\alpha$ (Mg,Si,Ca,Ti) and iron-peak
(Cr,Fe,Ni) elements. Also the Li and Al abundances and the total C+N+O content
are the same within the errors. 

On the other hand the light elements C,N,O,Na are different between the two
groups, with a significance of more than 4 $\sigma$. Also the carbon isotopic ratio
is different. This is shown in Fig.~\ref{f4}, were we plot [N/Fe]
vs. $^{12}$C/$^{13}$C. In this figure an anti-correlation appears, and the
difference in $^{12}$C/$^{13}$C between the two groups is more than 3
$\sigma$, although there is a substantial overlap of the two distributions due
in part to the measurement error.
This result is not unexpected because the N-rich population is supposed to be born from
material more chemically evolved with respect to the N-poor one. So it should have a lower
carbon isotopic ratio, as we find.

\begin{figure*}[ht]
\centering
\includegraphics[width=15cm]{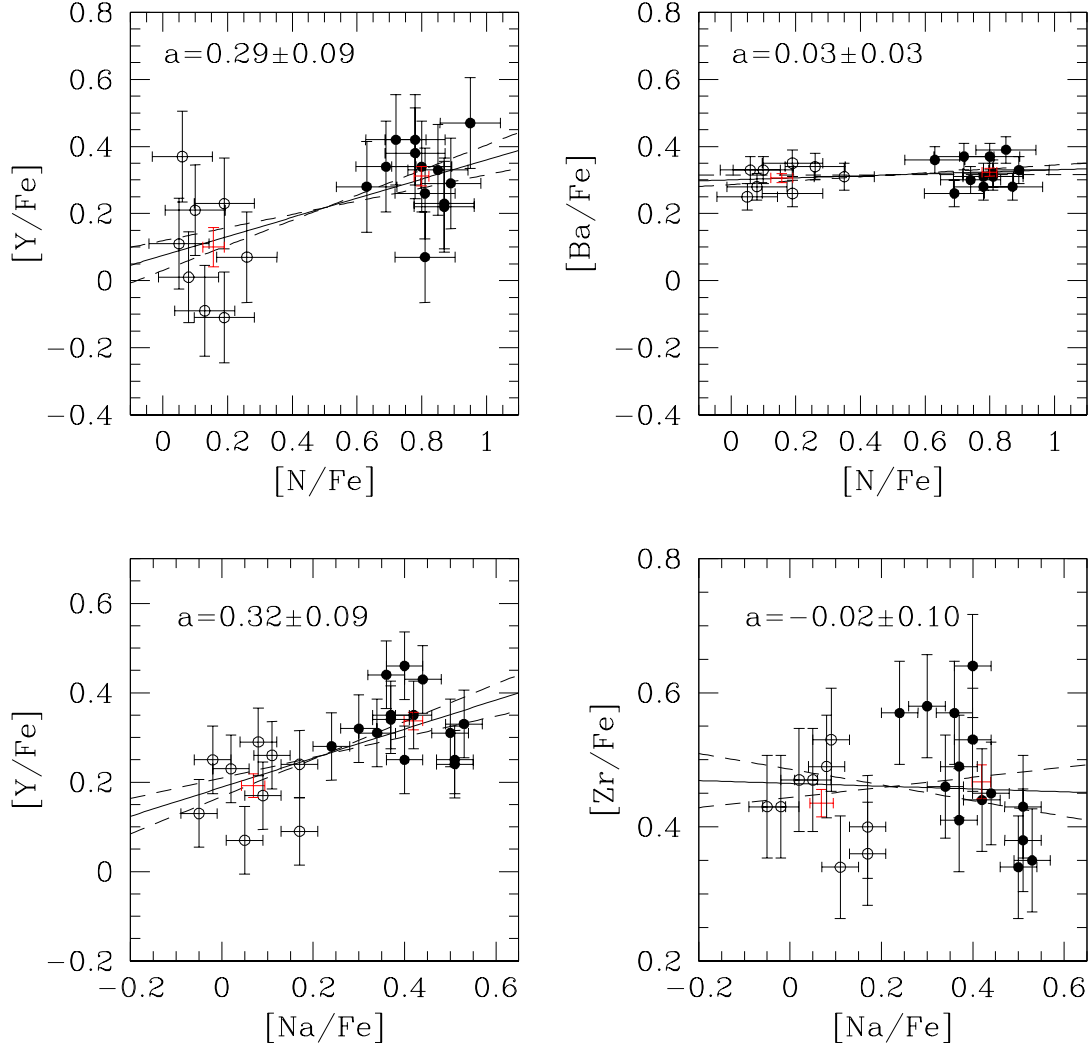}
\caption{Upper panels: [Y/Fe] and [Ba/Fe] as a function of [N/Fe] for our
GIRAFFE stars. Lower panels: [Y/Fe] and {Zr/Fe} as a function of [Na/Fe]
for out UVES stars. A straight line was fit to the data of each panel and
the slope with its error is reported.
Red crosses represent the mean value for each sub-population.}
\label{f5}
\end{figure*}

Fig.~\ref{f5} (upper panels) displays the abundances of the s-elements Y
and Ba for our GIRAFFE data. As in the case of the $\alpha$ and iron-peak
elements, the two sub-populations have the same Ba content. 
However the mean Y abundance is significantly different, at a level of more than 3
$\sigma$ (see Tab.~\ref{t2}). We fitted to the points a straight line and
calculated the slope and its error. Values are reported inside each panel.
The Ba slope is compatible with 0 (i.e. the same mean [Ba/Fe] value for the two
groups), while the Y slope is incompatible with 0 with a confidence of more than 3
$\sigma$ (i.e. two different mean [Y/Fe] values for the two groups, although
there is substantial overlap between the two distributions due in part to the
measurement error). We cannot rule out a trend, but we favor a bimodality.

Lower panels display the abundances of the s-elements Y and Zr for the
UVES data vs. [Na/Fe]. Again the mean Y content of Na-poor (N-poor) and
Na-rich (N-rich) stars is very different, at a level of more than 3 $\sigma$,
while they share the same Zr abundance.

The Y abundance from GIRAFFE and UVES observations deserves a further comment. For N-rich
stars the two databases give very good agreement (+0.31 vs +0.33 dex), well
within 1$\sigma$. Values for N-poor stars instead appear 
different ([Y/Fe]=+0.10$\pm$0.06 dex for GIRAFFE and [Y/Fe]=+0.19$\pm$0.03 dex for
UVES). However if we consider the errors, the difference of 0.09 dex is
significant at the level of 1.3 $\sigma$, too low to imply a real difference,
so we can safely attribute it to measurement errors.

The total error on Y (0.12 dex, see Tab.~\ref{t5}) due to atmospheric 
parameters and S/N is high. It is dominated by errors in gravity and
microturbulence, but also S/N gives a not-negligible contribution. However
this is a random internal error, and for this reason it is fully
included in the error of the mean Y abundance of the two groups of
stars that we use to calculate the significance in Tab.~\ref{t2}.

\begin{figure}[h]
\centering
\includegraphics[width=9cm]{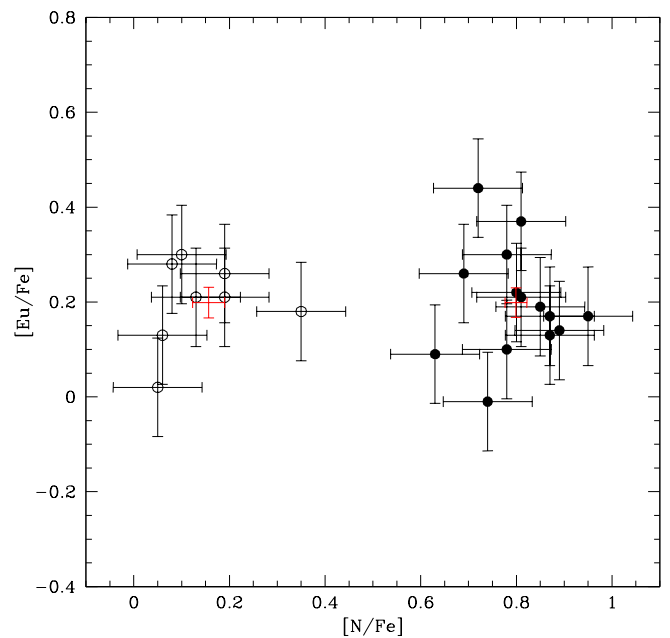}
\caption{[Eu/Fe] vs. [N/Fe] for our GIRAFFE stars.
Red crosses represent the mean value for each sub-population.}
\label{f6}
\end{figure}

Finally in Fig.~\ref{f6} we report [Eu/Fe] as a function of [N/Fe]. At odds with
the Y-abundance discussed before, Eu does not show any trend, confirming the
result reported in Tab.~\ref{t2}.

Thus, we find no significant difference between the two groups defined by their
N (or Na) abundance, in Li,C+N+O,Mg,Al,Si,Ca,Ti,Cr,Fe,Ni,Zr,Ba or Eu, but find
very significant differences in C,N,O,$^{12}$C/$^{13}$C,Na, and Y.

\subsection{Comparison with literature}

Firstly we compare our result with \citet{Ma08} (see Fig.~\ref{f3}).
Also in that case the authors find a bimodal Na-O anti-correlation.
They define a Na-poor (green points with [Na/Fe]$\leq$0.25) and a Na-rich
(green points with [Na/Fe]$>$0.25) population, which correspond to our N-poor
and N-rich respectively. Their Na-rich population have the same mean Na content as
our N-rich stars, while their Na-poor stars have a mean Na content that is
slightly higher than that of our N-poor population. This could be a residual of the
NLTE correction.
Here we show that the bimodality is extended also to C, N,$^{12}$C/$^{13}$C,
and Y. As \citet{Ma08}, we do not find bimodality in 
$\alpha$ (Mg,Si,Ca,Ti) or iron-peak elements(Cr,Fe,Ni), nor Ba.
In particular \citet{Ma08} Na-poor and Na-rich populations have the same [Ba/Fe]
within 1 $\sigma$, confirming our finding.
Thus the two populations have the same abundance as far as these
elements are concerned.

Another important paper is \citet{Iv99}.
These authors include also N and indirectly C in their results.
They find the same Na-O anticorrelation as we do (see their Fig. 14, upper
panel). They find also a relatively well-defined C-O correlation and N-O
anticorrelation as is implied also by our Fig.~\ref{f3}.
Their C+N+O content is constant within the errors for all the stars, 
but a bit higher that our result (8.24 vs. 8.16 respectively) in absolute
value. As in our case a bimodality is suggested by their measurement of the
strength of the CN band at 7874 \AA\ (see their Fig. 11), and partially visible 
in the N-O anticorrelation where a discontinuity at
log$\epsilon$(N)$\sim$7.70-7.80 is visible.

As suggested by the referee, we comment in more detail on the Al content. \citet{Ma08}
found a possible spread in Al, confirmed by \citet{Iv99}, where Al is
correlated with Na with a significance of more than 3$\sigma$.
We instead find that the Al content is the same for the two groups of stars, which
implies no Al-Na correlation.
A possible explanation could be the presence of some unrecognized molecular line
(CN?) that is blended with Al lines. This line
could preferentially affect colder stars, and both the \citet{Ma08} and
\citet{Iv99} stars are colder than ours. Because C and N have
different abundances for the two groups, we expect that this hypothetical line
has a different strength (assuming the same T$_{eff}$) if a star belongs to one
group or the other. This would cause a spurious correlation between Al and
N or Na also if the Al content is the same for the two groups.
This hypothesis is not totally unreasonable because the spectral ATLAS of
Arcturus \footnote{http://spectra.freeshell.org/spectroweb.html} shows that
the Al lines  are possibly blended with some CN line (i.e. the CN line at
$\lambda$= 6698.746 \AA), but the apparent lack of an increasing trend in Al
abundance with evolutionary phase in previous works could argue against it.

Another explanation could be the evolutionary state of \citet{Ma08} and
\citet{Iv99} targets. Their stars are all above the RGB-bump and so affected
by more evolutionary mixing. This is also proved by the $^{12}$C/$^{13}$C
value obtained by \citet{Iv99} that is of the order of 4-5. Our stars instead
have $^{12}$C/$^{13}$C$\sim$20, implying a less dramatic mixing.
A deeper mixing after the RGB-bump could alter significatively (and
differentially, with the N-rich group being the most affected) the primordial
Al content through products of the MgAl-cycle. Howver if nucleosynthesis
products of the MgAl-cycle are indeed mixed up at this evolutionary phase,
this would have consequences also for the Na abundances, as the NeNa-cycle
operates at lower temperatures \citep{Ch05}, and there are studies that argue
against such changes \citep{Gr00}.
A further discussion of this point is beyond the scope of this paper.
In any case, in the present paper we find that N-poor and N-rich stars have the
same Al content.

Our main results are independent of the absolute abundances, but we
want to add further comments on this point.
Comparison with absolute abundances published by \citet{Ma08}, \citet{Iv99}, 
\citet{Yo08}, and \citet{Yo08b} (the four most recent papers) are reported in
Tab.~\ref{t2}.
The last two papers are reported as one (Yo08) in the table and in the following
discussion because they are complementary.
For some elements the mean abundance of the cluster agrees well (difference
of 0.1 dex or less) with these four papers. This is true for C+N+O, Mg, Ti, Cr, and Fe.
$^{12}$C/$^{13}$C is higher in our case, as expected by the evolutionary
stage of our stars. Our Al, Si, and Ba content agrees well also with
\citet{Ma08}, but in this case the scatter is larger (between 0.1 to 0.3 dex) with respect to
\citet{Iv99} and Yo08. Our [O/Fe], [Na/Fe], and [Ni/Fe] match \citet{Ma08} and \citet{Iv99} within
0.1 dex, while the disagreement is worse (0.22, $\sim$0.2, and 0.13 dex respectivelly) with respect
to Yo08. On the other hand our [Ca/Fe] matches Yo08, but it is 0.13
and 0.15 dex higher then \citet{Ma08} and \citet{Iv99}.
The difference in C and Eu is a bit high
(0.22  and 0.15 dex respectively) with respect to \citet{Iv99}, but it is
large (0.37 dex) only in the case of N. Finally Yo08 obtained a Zr value 0.23 dex lower
than our, a Eu value 0.20 dex higher, while the
disagreement is large as far as [Y/Fe] is concerned ($\sim$0.4$\div$0.5 dex).

We further comment about Ba and Eu. We find [Ba/Fe]$\sim$+0.3, in agreement
with \citet{Ma08} who give [Ba/Fe]$\sim$+0.4. \citet{Iv99} gives a much
higher value ([Ba/Fe]$\sim$+0.6) while \citet{Gr86} gives [Ba/Fe]$\sim$+0.0.
Our value is in the middle of the literature values. We find [Ba/Eu]=+0.10,
lower than \citet{Iv99} but still higher than the solar system value and much
higher then field halo and globular cluster giants, where [Ba/Eu] is typically
negative with a range from -0.2 to -0.6 \citep{Iv99}. This result confirms
that M4 has a larger s- to r-process contribution than in the Sun, and also
supports the \citet{Iv99} suggestion that {\it the period of star formation and mass loss
that preceded the formation of the observed stars in M4 was long enough for
AGB stars to contribute their ejecta into the primordial ISM of the cluster}.

\section{Discussion}

As discussed in the introduction, all GCs studied in detail to date show some kind of spread
in their light-element (from C to Al) abundance. The amount of the
spread varies a lot from cluster to cluster. Some GCs, such as M22
\citep{Ma09}, show also a spread in $\alpha$ and iron-peak elements, but
this is an uncommon feature. So in this discussion we consider only
M4-like objects, i.e. those clusters having only a spread in light elements
(and possibly in s-elements). However M4 appears to be rather unique in that
the 'spread' is actually a bimodality.

The most natural explanation for this phenomenon is the self-pollution scenario,
where a first generation of stars is formed from primordial material. In the most
accepted model, this material is O-rich and Na-poor with respect to the
second generation that will form later. 
Then some class of stars (massive MS stars, either fast rotating or binaries,
or intermediate-mass AGB stars) of this first generation pollute the
interstellar material. This material (O-poor and Na-rich) is kept in the
cluster due to the strong gravitational field of the massive cluster, and it
gives rise to a new generation of stars. In this model the
O-rich/Na-poor stars are the oldest.
Also the abundance of other elements (including He or other light and s-process
elements) may differ in stars of the first and second generation.

Another scenario postulates instead that the primordial material the
first generation will form from is O-poor and Na-rich because of initial
pollution by SNeIa and AGB stars. Then SNeII of the first generation explode
polluting the residual material, and a second generation is formed, being
O-rich and Na-poor. The final result is the same, but in this case the
O-poor/Na-rich stars are the oldest.

Each one of these polluters has its own chemical signature as discussed in the
introduction. SNeII ejects material that is strongly $\alpha$-enhanced but only
slightly enhanced in iron-peak elements. The \citet{Mar09} scenario manages
to produce a second generation that has the same iron-peak element content
as the first generation because of the mix with primordial Fe-poor
material. Nothing is said in that paper about $\alpha$-elements with the
exception of O and Mg. However the first generation is poor in
$\alpha$-elements (Si,Ca,Ti) because it is formed from material polluted by SNeIa
and AGBs (that do not produce these elements or only in a negligible amount),
while the second generation must be rich in $\alpha$-elements
(Si,Ca,Ti) both because of the contamination by the SNeII of the first
generation, but also because of the mix with primordial material that
was pre-contaminated by previous SNeII.

The exact estimation of the difference in Si,Ca,Ti content is beyond the purpose
of this paper, but we can give a rough number. In \citet[Tab. 1]{Mar09} we
can see that the progenitor material of the first generation is enhanced by
$\sim$1 dex in its Fe content with respect to the primordial one, due to the SNIa
explosion. So the material the first generation will
form from is dominated by the products of the SNIa ejecta (plus the AGB
ejecta, that however do not affect Si,Ca,Ti content). The second generation
instead is formed from material whose abundance ratio is dominated by SNeII
ejecta. So we can estimate the difference in Si,Ca,Ti between the two
generations comparing the Si,Ca,Ti content of Galactic stars with [Fe/H]$\leq$-1.0
dex, whose abundance ratio is dominated by SNeII ejecta, and stars with
[Fe/H]$\sim$0.0, whose abundance ratio is dominated by SNeIa ejecta.
According to \citet[Fig. 8]{Po08}, the difference of the order of
$\Delta$[Si,Ca,Ti/Fe]$\sim$0.3. This is our reference value. 
Because SNeII produce also r-element (i.e. Eu),
the second generation should be also Eu-enhanced.

AGB stars are the main producers of s-elements, including both light-s such
as Y, Zr, and heavy-s like Ba, through the main component of the s-process.

Massive MS stars (M$>$15M$_{\odot}$, both as fast rotators or in binary systems)
produce only light s-elements (e.g. Y) through the weak-s process but not heavy
s-elements.

We now compare these predictions with our observational results.
N-poor and N-rich stars have the same Si,Ca,Ti content within a few hundredths
of a dex. They also share the same Eu abundance. According to these results
and comparing them with the Si,Ca,Ti content expected from \citet{Mar09}
scenario, SNeII are not viable candidates for the polluters responsible.

The two groups also have the same Ba content, suggesting that
AGB stars cannot be responsible for the pollution.
This statement is reinforced if we compare our results with theoretical
predictions. Considering both GIRAFFE and UVES databases, we have a difference
in [Y/Fe] for the two groups of stars of 0.18 dex. According to \citet{Bu01} or
the more recent paper by \citet{Ka10}, we would have expected to see an equal
(if not larger) difference in [Ba/Fe], assuming AGB stars as polluters. And we have
the observational counterpart of that. NGC1851 \citep{Vi10} hosts two 
distinct populations that differ in [Y/Fe] at the level of 0.11 dex. They have
a difference in [Ba/Fe] that is much larger, 0.41 dex. So for NGC1851 we can
postulate AGBs as the most probable polluters, while for M4 it is very unlikely.

\begin{figure}[h]
\centering
\includegraphics[width=9cm]{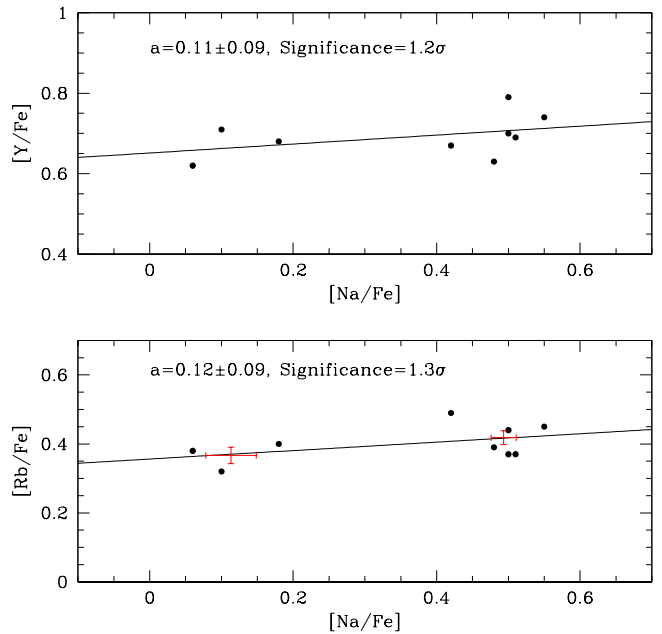}
\caption{[Y/Fe] vs. [Na/Fe] and [Rb/Fe] vs. [Na/Fe] for stars from \citet{Yo08b}.
Linear fits and the slopes with their significance (in units of $\sigma$) are
reported. For [Rb/Fe] vs. [Na/Fe] we plotted the mean abundances (red crosses) of the
two groups of Na-poor ([Na/Fe]$<$0.30 dex) and Na-rich ([Na/Fe]$>$0.30 dex) stars.}
\label{f7}
\end{figure}

We are left with massive MS stars. These objects should pollute the interstellar
material with light s-elements (besides C,N,O,Na,Al)  produced through the weak s-process. We
found that the two populations differ in their Y content at a level of more
than 3 $\sigma$, but they have the same Zr content. This results fits
with the theoretical scenario \citep{Ra93,Tr04} that says that the weak
s-component is responsible for a major contribution to the s-process nuclides
up to A$\sim$90 with a peak for A=80 ($^{80}$Kr). The last nuclide of the
chain is uncertain due to the many theoretical uncertainties,
but it is not unreasonable (and it is compatible with the models) to postulate that,
according to our results, {\it the last nuclide produced in a significant
amount is $^{89}$Y}, leaving the next nuclide, $^{90}$Zr, and heavier
s-elements, relatively scarce.
Unfortunately we cannot measure s-elements lighter then Y (i.e. Rb or Sr) in
order to further investigate our statement. However we can obtain
an important confirmation from \citet{Yo08b}. These authors estimated Na, Rb,
Y, and Pb for a sample of targets, and their results are reported in Fig.~\ref{f7} for those stars with
all three elements measured. First of all also in
this case Na appears to be bimodal, with a gap between 0.2$<$[Na/Fe]$<$0.4.
But, most important, both Y and Rb appear to have a positive trend with
Na (or N), exactly as we have found (but only for Y). This would confirm that
light s-elements like Y or lighter have a different mean abundance in the two
populations of M4. The trends, taken separately, have a significance of only 1.2$\div$1.3 $\sigma$, as shown in
Fig.~\ref{f7}, but if we compare the mean Rb content of the two groups
(red crosses in Fig.~\ref{f7}, lower panel) that are of +0.37$\pm$0.02 and +0.42$\pm$0.02
respectively, we have a significance of 1.8 $\sigma$.
We obtain an even stronger confirmation if we consider the two trends
together and apply the following ab absurdum argument.
Let's assume that our previous result is wrong and that the two populations
have the same Rb and Y content. With this hypothesis and using the
Kolmogorov-Smirnov test, the probability of having
a Y vs. Na data distribution like that one in Fig.~\ref{f7} is 17\%.
For the Rb vs. Na data distribution we have a probability of 9\%. These two values
match with the significance at 1.2$\div$1.3 $\sigma$ found above.
Under the previous hypothesis the relations visible in Fig.~\ref{f7} would
be due only to measurement errors, so they would be independent and 
we can calculate the probability of having both relations we see in
Fig.~\ref{f7} by multiplying the two individual values.
The final result is only 1.5\%.
So the probability of the initial hypothesis to be true is only 1.5\%.
This means that \citet{Yo08b} confirms our result of a bimodality in light
s-elements with a confidence level of 98.5\%.
Another hint comes from Pb. This is a pure s-element, produced only in
AGB stars \citep{Tr04}. The mean [Pb/Fe] abundances of the two Na groups in
\citet{Yo08b} are +0.26$\pm$0.05 and +0.31$\pm$0.02 respectivelly.
In this case the significance is 0.8 $\sigma$. This further rules out AGB
stars as candidate polluters, in agreement with our statement based on Ba.
Because of this and because we have double-checked with two different spectrographs and
two different lines the Y-bimodality, we are led to the conclusion that
the best candidates for the Y-enrichment for the second generation in M4 is
the weak s-process, which implies that massive (M$>$15 M$_{\odot}$) MS stars
are the best candidates for the self-pollution scenario.

A further conclusion can be obtained from Mg and Al. These elements are
processed in the Mg-Al cycle where Al is produced
at the expense of Mg. The fact that we find the same Mg and Al abundance for
N-poor and N-rich stars implies that the Mg-Al cycle did not activate, at least in the
ejected material responsible for the pollution. This implies an upper limit for the
temperature: the material burns at a temperature lower than 50$\times$10$^6$ K
\citep{De07}. This implies also that the massive stars responsible for the pollution
must have been less massive than 60 M$_{\odot}$. Another hint comes from the
difference in the mean [O/Na] value of the two groups. In our database it is 
of the order of 0.7 dex, while \citet{Ma08} give 0.5 dex, so we assume a mean value
of 0.6 dex. \citet{De07} (see their Fig. 10) give the extension of the [O/Na]
distribution as a function of the mass of the polluter. 0.6 dex implies a
mass higher then 20 M$_{\odot}$ but significantly lower than 30 M$_{\odot}$. A value
between 20 and 30 M$_{\odot}$ seems reasonable.

Our conclusion is that the best candidate in order to explain the
abundance spread in M4 are massive MS stars (fast rotators or binaries) with
masses of the order of 20$\div$30 M$_{\odot}$.
Interestingly enough, there is another cluster with the same chemical
signature as M4, that is NGC~6397, recently investigated by \citet{Li11}.
In that paper the authors find a bimodality in Na for the cluster as in M4,
but also a difference in [Y/Fe] between the two sub-populations at a level of
almost 3$\sigma$. As in our case the content of the other neutron-captured elements
(Zr,Ba,Ce,Nd,Eu), all heavier than Y, is the same within $\sim$1$\sigma$. They
say that the [Y/Fe] differecence is 'too small to be convincing' (citation),
but in the view of our result it could be real. Based on the
abundance pattern, they support the MS massive star scenario too.
These findings suggest also the hypothesis that a bimodality in
light-elements (C,N,O,Na) could be a normal feature of the MS massive star based
self-pollution scenario.

We underline the fact that our conclusion is valid for M4 (and possibly
NGC~6397). Other clusters show a Mg-Al anticorrelation or light and heavy s-element spread
(i.e. NGC~1851, \citealt{Vi10}), so AGB stars can be involved.

Finally some clusters like $\omega$ Centauri or M22 have a spread in Ca and Fe, so SNeII must
also be at work. The picture is that each cluster may have its own peculiar process of
formation within the pollution scenario, and that the polluters can be
different.

A peculiarity of M4 is its bimodal distribution, at odds with many other
clusters where the Na-O anticorrelation is continuous \citep{Ca09}. This
implies that the clusters had a first star forming burst followed by a 
quiescent period, where processed material started to flow and mix with the
primordial gas. During this period no star was formed and the gas had time
to homogenize, otherwise we would have observed a continuous anticorrelation as
in the other clusters.
After that, a second burst happened with the formation of the second generation.
The exact time scale of this process is unknown. However it must have been longer
than the evolutionary time of very massive stars (some Myrs), but shorter
than the evolutionary time of AGB stars ($\geq$40 Myrs, \citealt{Ve09}).
For the 20$\div$30 M$_{\odot}$ star polluters postulated above, the
evolutionary timescale is of the order of $\sim$10$\div$30 Myrs.

\section{Conclusions}

In this paper we analyzed a sample of 23 stars belonging to the RGB of M4
(NGC~6121) and observed with the GIRAFFE@FLAMES spectrograph. Our targets are
located below the RGB bump. We complemented our study by analyzing a subsample of
UVES@FLAMES spectra of \citet{Ma08}. We estimated abundances for many key
elements: Li,C,N,O,$^{12}$C/$^{13}$C,Na,Mg,Al,Si,Ca,Ti,Cr,Fe,Ni,Zr,Y,Ba,Eu.
Our aim was to look for some hint in order to help solve the problem related to the
self-pollution scenario in GCs. According to this scenario, a cluster
experiences an extended  period of star formation, where the younger
populations were born from an interstellar medium polluted by products of the
CNO, NeNa, and MgAl cycles coming from massive stars of the former generation.
Postulated polluters include: massive MS stars (M$>$15 M$_{\odot}$, fast
rotating or binaries), intermediate mass AGB stars (4$<$M$<$7 M$_{\odot}$), or
SNeII. Each of these candidates has its own unique chemical signature.
We confirm the presence of a bimodal population, where the two groups of stars
can easily be separated by their N content. One group (presumably the 
oldest generation) is N-poor, while the other (the younger generation) is
N-rich. N-poor and N-rich stars have
significantly different C,N,O,$^{12}$C/$^{13}$C,Na, and Y content, but share the same
Li,C+N+O,Mg,Al,Si,Ca,Ti,Cr,Fe,Ni,Zr,Ba, and Eu abundances. This rules out SNeII because
they would produce a second generation $\alpha$-enhanced with respect to the
first one (and Eu enhanced). AGB stars are also excluded
because they should produce a Zr and Ba-enhanced (in addition to Y-enhanced)
second generation through the main s-process.
We are left with massive MS stars. These objects are able to produce the
difference in light elements we observe but, most important, they can produce
also light s-elements like Y through the weak s-process. The Y-enhancement
of the second generation is the most interesting result we found. 
This scenario is supported also by \citet{Yo08b}. Based in this paper we
found that the two M4 groups have different Rb content (that confirms massive
MS stars as polluters) but the same Pb content (that further excludes AGB stars
as polluters).
The lack of Mg/Al enhancement and the extension of [O/Na] ratio points toward
massive stars with 20$<$M$<$30 M$_{\odot}$ as the most likely polluters.

Our conclusion is that massive stars in the range 20$<$M$<$30 M$_{\odot}$ are
responsible for the bimodal population in M4. The time scale for the formation 
of the clusters is $\sim$10$\div$30 Myrs, with two well separated bursts that
generated a bimodal population.
Other cluster have different chemical characteristics, so other kinds of
polluters and star formations histories are required.

\begin{acknowledgements}
S.V. and D.G.gratefully acknowledge support from the Chilean
{\sl Centro de Astrof\'\i sica} FONDAP No. 15010003
and the Chilean Centro de Excelencia en Astrof\'\i sica
y Tecnolog\'\i as Afines (CATA). S.V. and D.G. gratefully acknowledge
also the referee that helped clarify and strengthen a number of important
points.
\end{acknowledgements}

\clearpage

\begin{table*}
\caption{Individual abundances (light, $\alpha$, iron-peak elements) of the observed GIRAFFE stars.}            
\label{t3}      
\centering
\begin{tabular}{lcccccccccccccc}        
\hline\hline
\tiny{ID} & \tiny{log$\epsilon$(Li)} & \tiny{[C/Fe]} & \tiny{[N/Fe]} & \tiny{[O/Fe]} & \tiny{[Na/Fe]} & \tiny{[Mg/Fe]} & \tiny{[Al/Fe]} & \tiny{[Si/Fe]} & \tiny{[Ca/Fe]} & \tiny{[Ti/Fe]} & \tiny{[Cr/Fe]} & \tiny{[Fe/H]} & \tiny{[Ni/Fe]} & \tiny{$^{12}$C/$^{13}$C}\\
\hline             
\tiny{28590} & 1.02 & -0.42 & 0.89 & 0.26 &  0.48 & 0.48 & 0.66 & 0.38 &  -   & 0.26 &   -   & -1.21 & -0.05 & 18\\
\tiny{33584} & 0.94 & -0.21 & 0.35 & 0.34 &  0.01 & 0.54 & 0.64 & 0.45 & 0.41 & 0.31 &   -   & -1.14 & -0.04 & 20\\
\tiny{36820} & 0.77 & -0.27 & 0.78 & 0.34 &  0.45 & 0.49 & 0.64 & 0.49 & 0.42 & 0.32 &  0.10 & -1.25 & -0.02 & 20\\
\tiny{37614} & 1.08 & -0.18 & 0.26 & 0.43 & -0.08 &  -   & 0.62 & 0.50 & 0.36 & 0.34 &  0.02 & -1.16 &  0.04 & 22\\
\tiny{39100} & 0.97 & -0.19 & 0.05 & 0.37 &  0.03 & 0.42 & 0.44 & 0.47 & 0.40 & 0.36 & -0.03 & -1.08 &  0.00 & 25\\
\tiny{40197} & 0.97 & -0.27 & 0.78 & 0.40 &  0.35 & 0.51 & 0.51 & 0.51 & 0.42 & 0.28 & -0.06 & -1.14 & -0.03 & 22\\
\tiny{41863} & 0.91 & -0.38 & 0.87 & 0.26 &  0.57 & 0.51 & 0.61 & 0.48 &  -   & 0.23 &  0.18 & -1.19 & -0.03 & 21\\
\tiny{42561} & 0.96 & -0.40 & 0.95 & 0.28 &  0.37 &  -   & 0.62 & 0.39 & 0.41 & 0.31 & -0.06 & -1.20 &  0.04 & 13\\
\tiny{43020} & 1.05 & -0.19 & 0.63 & 0.45 &  0.23 & 0.47 & 0.49 & 0.51 & 0.50 & 0.28 &  0.01 & -1.12 & -0.04 & 22\\
\tiny{43085} & 1.01 & -0.32 & 0.80 & 0.31 &  0.40 & 0.47 & 0.50 & 0.42 & 0.33 & 0.27 & -0.05 & -1.08 &  0.02 & 19\\
\tiny{43494} & 0.93 & -0.18 & 0.13 & 0.48 & -0.11 & 0.49 & 0.31 & 0.43 & 0.44 & 0.31 &  0.00 & -1.14 & -0.05 & 22\\
\tiny{43663} & 1.00 & -0.28 & 0.69 & 0.30 &  0.31 & 0.36 & 0.43 & 0.45 & 0.33 & 0.30 &  0.06 & -1.11 & -0.07 & 19\\
\tiny{45171} & 0.80 & -0.13 & 0.06 & 0.53 & -0.09 & 0.32 & 0.44 & 0.40 & 0.44 & 0.38 & -0.09 & -1.11 &  0.00 & 21\\
\tiny{45200} & 0.94 & -0.42 & 0.87 & 0.14 &  0.55 & 0.51 & 0.57 & 0.46 & 0.39 & 0.36 &  0.07 & -1.12 &  0.00 & 12\\
\tiny{46201} & 0.94 & -0.40 & 0.81 & 0.10 &  0.41 & 0.54 & 0.55 & 0.31 & 0.44 & 0.34 & -0.04 & -1.17 & -0.04 & 20\\
\tiny{47596} & 1.00 & -0.15 & 0.08 & 0.48 & -0.03 & 0.44 & 0.54 & 0.35 & 0.42 & 0.31 &  0.04 & -1.16 & -0.02 & 22\\
\tiny{48499} & 1.17 & -0.15 & 0.10 & 0.53 &  0.03 & 0.47 & 0.60 & 0.54 & 0.47 & 0.39 & -0.03 & -1.15 &  0.04 & 25\\
\tiny{49381} & 0.95 & -0.27 & 0.19 & 0.35 &  0.02 & 0.59 & 0.47 & 0.38 & 0.45 & 0.42 &  0.13 & -1.19 &  0.01 & 18\\
\tiny{50032} & 1.02 & -0.45 & 0.85 & 0.11 &  0.32 & 0.43 & 0.50 & 0.43 & 0.33 & 0.33 &   -   & -1.04 &  0.01 & 12\\
\tiny{53602} & 1.14 & -0.36 & 0.72 & 0.23 &  0.36 & 0.36 & 0.39 & 0.37 & 0.45 & 0.33 & -0.11 & -1.03 & -0.01 & 18\\
\tiny{67553} & 0.90 & -0.30 & 0.19 & 0.30 &  0.14 & 0.39 & 0.49 & 0.35 & 0.39 & 0.29 & -0.05 & -1.16 &  0.04 & 20\\
\tiny{ 8460} & 0.99 & -0.36 & 0.74 & 0.12 &  0.40 & 0.50 & 0.50 & 0.37 & 0.41 & 0.40 &   -   & -1.16 & -0.04 & 14\\
\tiny{  907} & 0.81 & -0.48 & 0.81 & 0.21 &  0.42 & 0.56 & 0.46 & 0.37 & 0.42 & 0.31 &  0.03 & -1.20 & -0.03 & 14\\
\hline                                   
\multicolumn{15}{c}{Sun:our linelist}\\
\hline
             &  -   &  8.49 & 7.95 & 8.83 &  6.32 & 7.56 & 6.43 & 7.61 & 6.39 & 4.94 &  5.63 &  7.50 &  6.26 & - \\
\hline
\multicolumn{15}{c}{Sun:\citet{Gr98}}\\
\hline
             &  -   &  8.52 & 7.92 & 8.83 &  6.33 & 7.58 & 6.47 & 7.55 & 6.36 & 5.02 &  5.67 &  7.50 &  6.25 & - \\
\hline
\end{tabular}
\end{table*}

\clearpage

\begin{table*}
\caption{Individual abundances (s and r elements) of the observed GIRAFFE and
UVES stars. For UVES also Na abundances from \citet{Ma08} are reported.}            
\label{t4}      
\centering                          
\begin{tabular}{lccc}        
\hline
\multicolumn{4}{c}{GIRAFFE}\\
\hline                 
\tiny{ID}    & \tiny{[Y/Fe]}&\tiny{[Ba/Fe]}&\tiny{[Eu/Fe]}\\
\hline             
\tiny{28590} &  0.29 & 0.33 &  0.14 \\
\tiny{33584} &   -   & 0.31 &  0.18 \\
\tiny{36820} &  0.38 & 0.31 &  0.30 \\
\tiny{37614} &  0.07 & 0.34 &   -   \\
\tiny{39100} &  0.11 & 0.25 &  0.02 \\
\tiny{40197} &  0.42 & 0.28 &  0.10 \\
\tiny{41863} &  0.23 & 0.28 &  0.17 \\
\tiny{42561} &  0.47 &  -   &  0.17 \\
\tiny{43020} &  0.28 & 0.36 &  0.09 \\
\tiny{43085} &  0.34 & 0.37 &  0.22 \\
\tiny{43494} & -0.09 &  -   &  0.21 \\
\tiny{43663} &  0.34 & 0.26 &  0.26 \\
\tiny{45171} &  0.37 & 0.33 &  0.13 \\
\tiny{45200} &  0.22 &  -   &  0.13 \\
\tiny{46201} &  0.07 & 0.32 &  0.21 \\
\tiny{47596} &  0.01 & 0.28 &  0.28 \\
\tiny{48499} &  0.21 & 0.33 &  0.30 \\
\tiny{49381} & -0.11 & 0.26 &  0.21 \\
\tiny{50032} &  0.33 & 0.39 &  0.19 \\
\tiny{53602} &  0.42 & 0.37 &  0.44 \\
\tiny{67553} &  0.23 & 0.35 &  0.26 \\
\tiny{ 8460} &   -   & 0.30 & -0.01 \\
\tiny{  907} &  0.26 & 0.31 &  0.37 \\
\hline                                
\multicolumn{4}{c}{Sun:our linelist}\\
\hline
      &  2.25 & 2.34 & 0.52\\
\hline                                
\multicolumn{4}{c}{Sun:\citet{Gr98}}\\
\hline
      &  2.24 & 2.13 & 0.51\\
\hline
\multicolumn{4}{c}{\ }\\
\hline
\multicolumn{4}{c}{UVES}\\
\hline
\tiny{ID}    & \tiny{[Na/Fe]}&\tiny{[Y/Fe]}&\tiny{[Zr/Fe]}\\
\hline
\tiny{19925} &  0.51 & 0.43 & 0.24\\
\tiny{20766} &  0.53 & 0.35 & 0.33\\
\tiny{21191} &  0.51 & 0.38 & 0.25\\
\tiny{21728} &  0.37 & 0.41 & 0.35\\
\tiny{22089} &  0.50 & 0.34 & 0.31\\
\tiny{24590} &  0.30 & 0.58 & 0.32\\
\tiny{25709} &  0.34 & 0.46 & 0.31\\
\tiny{26794} &  0.36 & 0.57 & 0.44\\
\tiny{27448} &  0.11 & 0.34 & 0.26\\
\tiny{28103} &  0.17 & 0.40 & 0.09\\
\tiny{28356} &  0.37 & 0.49 & 0.34\\
\tiny{28797} &  0.44 & 0.45 & 0.43\\
\tiny{28847} &  0.08 & 0.49 & 0.29\\
\tiny{28977} &  0.40 & 0.64 & 0.46\\
\tiny{29027} &  0.02 & 0.47 & 0.23\\
\tiny{29065} &  0.17 & 0.36 & 0.24\\
\tiny{29222} &  0.24 & 0.57 & 0.28\\
\tiny{29272} &  0.05 & 0.47 & 0.07\\
\tiny{29282} &  0.42 & 0.44 & 0.35\\
\tiny{29545} & -0.02 & 0.43 & 0.25\\
\tiny{29598} &  0.40 & 0.53 & 0.25\\
\tiny{29848} &  0.09 & 0.53 & 0.17\\
\tiny{30209} & -0.05 & 0.43 & 0.13\\
\hline
\multicolumn{4}{c}{Sun:our linelist}\\
\hline
             &  6.32 & 2.25 & 2.56\\
\hline
\multicolumn{4}{c}{Sun:\citet{Gr98}}\\
\hline
             &  6.33 & 2.24 & 2.60\\
\hline
\end{tabular}
\end{table*}

\clearpage

\begin{table*}
\caption{Estimated errors on abundances due to errors on atmospheric
parameters and to spectral noise compared with the observed errors}            
\label{t5}      
\centering
\begin{tabular}{lcccccccc}        
\hline\hline
ID & $\Delta$T$_{\rm eff}$=+50 K  & $\Delta$log(g)=+0.10 & $\Delta$[Fe/H]=+0.05 &\
 $\Delta$v$_{\rm t}$=+0.10 km/s & S/N & $\sigma_{\rm tot}$ & $\sigma_{\rm obs}$\\
\hline       
$\Delta$(log$\epsilon$(Li)) & +0.05 &  0.00 &  0.00 &  0.00 & 0.07 & 0.09 & 0.10\\
$\Delta$([C/Fe])            & -0.02 &  0.01 &  0.03 &  0.03 & 0.05 & 0.07 & 0.07\\
$\Delta$([N/Fe])            &  0.05 &  0.01 &  0.02 &  0.03 & 0.05 & 0.08 & 0.09\\
$\Delta$([O/Fe])            & -0.04 &  0.08 &  0.01 &  0.03 & 0.06 & 0.11 & 0.10\\
$\Delta$([Na/Fe])           & -0.02 &  0.00 &  0.00 &  0.02 & 0.08 & 0.08 & 0.08\\
$\Delta$([Mg/Fe])           & -0.02 & -0.02 &  0.01 &  0.00 & 0.08 & 0.09 & 0.07\\
$\Delta$([Al/Fe])           & -0.02 &  0.01 &  0.00 &  0.03 & 0.07 & 0.08 & 0.09\\
$\Delta$([Si/Fe])           & -0.04 &  0.02 &  0.01 &  0.02 & 0.06 & 0.08 & 0.06\\
$\Delta$([Ca/Fe])           &  0.00 &  0.00 &  0.00 &  0.01 & 0.04 & 0.04 & 0.04\\
$\Delta$([Ti/Fe])           &  0.02 &  0.00 &  0.00 &  0.01 & 0.03 & 0.04 & 0.04\\
$\Delta$([Cr/Fe])           &  0.00 &  0.00 &  0.00 &  0.02 & 0.07 & 0.07 & 0.08\\
$\Delta$([Fe/H])            & +0.05 & -0.01 & -0.01 & -0.03 & 0.01 & 0.06 & 0.05\\
$\Delta$([Ni/Fe])           & -0.01 &  0.01 &  0.00 &  0.01 & 0.03 & 0.03 & 0.03\\
$\Delta$([Y/Fe])            & -0.03 &  0.06 &  0.01 & -0.08 & 0.06 & 0.12 & 0.14\\
$\Delta$([Ba/Fe])           & -0.02 &  0.03 &  0.03 & -0.02 & 0.04 & 0.06 & 0.04\\
$\Delta$([Eu/Fe])           & -0.05 &  0.04 &  0.01 &  0.03 & 0.07 & 0.10 & 0.10\\
$\Delta$($^{12}$C/$^{13}$C) &   0   &   0   &   0   &   0   & 3    &   3  &  4  \\
\hline
\end{tabular}
\end{table*}

\begin{table*}
\caption{Equivalent Widths. Table \ref{t6} is published in its entirety in the
electronic edition of the Journal.}            
\label{t6}      
\centering
\begin{tabular}{lcccc}        
\hline\hline
Wavelength(\AA) & Element & E.P.(eV) & log(gf) & \#28590\\
\hline       
5711.083 & 12.0 & 4.34 & -1.67 &  86.4\\
6696.014 & 13.0 & 3.14 & -1.56 &  34.7\\
6698.663 & 13.0 & 3.14 & -1.83 &  14.3\\
5645.603 & 14.0 & 4.93 & -2.12 &  23.6\\
5690.419 & 14.0 & 4.93 & -1.84 &  35.0\\
5793.066 & 14.0 & 4.93 & -2.02 &  28.4\\
6125.014 & 14.0 & 5.61 & -1.58 &  14.3\\
6145.010 & 14.0 & 5.61 & -1.45 &  20.7\\
6244.465 & 14.0 & 5.62 & -1.34 &  23.9\\
6161.287 & 20.0 & 2.52 & -1.29 &  54.3\\
6162.170 & 20.0 & 1.90 &  0.46 & 170.9\\
6166.429 & 20.0 & 2.52 & -1.14 &  53.2\\
6126.214 & 22.0 & 1.07 & -1.36 &  26.8\\
6258.098 & 22.0 & 1.44 & -0.34 &  49.7\\
6261.094 & 22.0 & 1.43 & -0.44 &  42.4\\
\hline
\end{tabular}
\end{table*}

\end{document}